\documentclass[aps,pra,preprint,tightenlines,groupedaddress,showpacs]{revtex4}
\usepackage{graphicx}
\begin{document}
\preprint{DFF 406--5--03}
\title
{\bf Fermion pairing in Bose--Fermi mixtures}
\author{ F. Matera }
\email{matera@fi.infn.it}
\affiliation{
{\small\it Dipartimento di Fisica, Universit\`a degli Studi di Firenze, and}
{\small\it Istituto Nazionale di Fisica Nucleare, Sezione di
Firenze,}\\
{\small\it Via G. Sansone 1, I-50019, Sesto Fiorentino, (Firenze), Italy}}
\begin{abstract}
An effective interaction between fermions in a Bose--Fermi mixture is 
derived. It is induced by density fluctuations of the bosonic background. 
The contributions from states containing both one and two virtual phonons 
are taken into account self--consistently. The time dependence of the 
effective interaction has been removed by assuming that the velocity 
of the fermions at the Fermi surface is much larger than the sound 
velocity in the Bose gas. This assumption is more appropriate for the actual 
experimental situations than the usual approximation of neglecting 
retardation effects. The effective interaction turns out to be  
attractive and, as a consequence, can give rise to a  
superconducting phase in the Fermi component of 
the mixture. The fermions are considered in only one magnetic state, so 
that pairing can be effective only in odd--$l$ channels. It has been 
found that the onset of the superconducting phase can occur at 
temperatures (\,$>100\,{\rm nK}$\,) of the same order of magnitude 
as the Fermi temperature (\,$\sim 300\,{\rm nK}$\,), and the energy 
gap in the excitation spectrum is a small fraction (\,$\sim 1\%$\,) 
of the Fermi energy.
\end{abstract}
\maketitle

\section{Introduction}
Effective fermion--fermion interactions induced by the exchange of virtual  
phonons in fermion--boson mixtures were already studied in the 1960s,  
in connection with dilute solutions of $^3{\rm He}$ in superfluid 
$^4{\rm He}$, see e.g. Ref. \cite{Bar67}. 
Recently, a renewed interest in this subject has 
arisen, following the actual availability of trapped atomic gases  
at a temperature sufficiently low to exhibit quantum properties. 
In particular, physical situations are reproduced, in which a highly 
condensed Bose gas is mixed with an almost degenerate Fermi gas 
and the interaction between these two gases can be studied at various 
values of temperature and density \cite{Tru01,Sch01,Had02,Roa02}. 
On the theoretical side, particular attention 
has been devoted to the possibility of the onset of a supercoducting phase  
in the Fermi component of the mixture \cite{Bij00,Hei00,Efr02,Viv02}. 
For the fermion gases used in experiments the bare interatomic 
potential is repulsive. Then, the necessary attractive 
interaction between two fermions should be 
provided by the exchange of virtual phonons, through a 
mechanism similar to that responsible for the superconductivity   
phenomenon in metals. Investigations of the physical conditions for a 
a Cooper instability in the $s$--wave and $p$--wave channels have 
recently been performed \cite{Bij00,Hei00,Efr02,Viv02}. \par
For a bare fermion--boson interaction with zero--range, the induced 
fermion--fermion interaction is essentially proportional to the 
correlator of density fluctuations of the bosonic background. 
In the references \cite{Bar67,Bij00,Hei00,Efr02,Viv02}, the density 
fluctuations have been calculated taking into account only single--
phonon intermediate states. In the present paper the contribution 
from two--phonon intermediate states is also included. This does not 
simply amount to an improvement of the approximation scheme, because 
two--phonon and one--phonon states occur on the same basis  
in the evaluation of the effective fermion--fermion 
interaction, at lowest order in the fermion--boson coupling 
constant. Moreover, when the fractions of condensed and 
non--condensed components of the boson gas are comparable, 
the one--phonon and two--phonon contributions 
turn out to be of the same order of magnitude. 
\par
The induced interaction between fermions is given by 
the boson density--density correlator, evaluated for the time 
interval between two successive fermion--boson interactions. 
In order to simplify calculations, different approximations    
can be introduced to eliminate this time dependence.   
In Refs. \cite{Bar67,Bij00,Hei00,Efr02,Viv02} retardation effects have 
been neglected, always by setting the frequency to zero in the time--Fourier 
transform of the correlator. For a degenerate Fermi gas, this 
approximation can be justified if the Fermi velocity is 
much lower than the sound velocity in the Bose gas. 
In the present approach the opposite approximation is used: 
the correlator is taken at equal times, i.e. the induced interaction 
between two fermions is mediated by the static correlation function for 
boson density fluctuations. This approximation is more appropriate  
for the physical conditions obtained in the mixtures of 
trapped Bose--Fermi atoms produced in the laboratory 
\cite{Tru01,Sch01,Had02,Roa02}. For these mixtures 
the density of the degenerate Fermi gas is a 
relevant fraction of the density of the Bose gas. \par
The aim of the present work is to assess the possible 
occurrence of a superconducting phase in the Fermi component of 
an ultracold mixture of Bose--Fermi gases, for values of density and 
temperature obtained in actual experiments, 
and to give an estimate of the gap in the spectrum of excitation 
energies. Although all the experiments have been 
performed with harmonic--oscillator traps, a spatially 
homogeneous system is considered. In this case,   
calculations are considerably simplified and   
compact analytical expressions can be derived. 
\par
At the densities considered in the experiments, for the two species  
of gases we can assume that the interatomic interactions   
have a short range and mainly occur in the $s$--wave channel. 
Moreover, according to the experimental situations, both the Fermi  
and Bose atoms are considered in only one internal state. 
In this case, the Pauli principle allows us to neglect the bare 
interaction between fermions, at least in a mean--field 
treatment. In fact, in the zero--range limit 
for the interaction, the contributions from the 
direct ( Hartree ) and  exchange ( Fock ) terms of the mean--field 
cancel each--other exactly, while short range  correlations 
should not play an important role for a sufficiently dilute  
Fermi gas, i.e. for $a_{FF}^3\,n_F<<1$ where $a_{FF}$ is the 
scattering length and $n_F$ is the fermion density. Concerning 
the Cooper condensation, it can occur only for odd values of 
the orbital angular momentum for a spin--aligned couple of fermions, 
so that there is no contribution from a $s$--wave interaction. 
\par
Experiments have been performed with gases, which show  
repulsive boson--boson interactions \cite{Sch01,Roa02},  
this ensures the stability of the mixture towards separation  
of the two species of gases allowing us to study mixtures with a 
sufficiently high number of atoms. Moreover, the interaction between 
a fermion and a boson in the mixture $^{40}{\rm K}$--$^{87}{\rm Rb}$ 
of the Florence experiment \cite{Roa02} is strongly 
attractive, thus the induced fermion--fermion interaction 
can be very effective to bind fermions in correlated pairs. 
The calculations are performed on a general ground, however, 
when more phenomenological aspects are concerned, they mostly refer 
to the physical environment of Ref. \cite{Roa02}. 
\section{Formalism}
In this section the effective interaction between two 
fermions, mediated by density fluctuations of the Bose gas, 
and the equation for the energy gap in the excitation spectrum 
of the Fermi gas are derived. The calculations are performed by means of 
functional methods thoroughly, since both the 
properties of the Bose condensate and the induced 
fermion--fermion interaction can be directly obtained 
in the same scheme (~for a introduction to functional methods 
see e.g. Ref. \cite{Negele,Sto99}~). 
\subsection{Effective interaction}
The starting point is the functional--integral expression for the 
grand--canonical partition function of the Bose--Fermi mixture, 
that reads 
\begin{equation}
Z= \int\,{\cal D}(\phi, \phi^*,\psi,\psi^*)\,{\rm exp}\Big[-
\Big(S_B(\phi, \phi^*)+S_F(\psi,\psi^*)+S_I(\phi, \phi^*,\psi,\psi^*)\Big)
\Big]\, .
\label{grpart}
\end{equation}
The integration is performed over the complex field 
$\phi({\bf x},\tau)$ (\,bosons\,) and the Grassman field 
$\psi({\bf x},\tau)$ (\,fermions\,), which are respectively 
periodic and anti--periodic in the imaginary--time 
interval $(0,\beta=1/T)$ (~units such that $\hbar=~c=~k_B=1$ are used~). 
Since we deal with spin--polarized particles, 
$\phi({\bf x},\tau)$ and $\psi({\bf x},\tau)$ are one--component  
fields. The action functional has been written as the sum of 
three terms: the action for the boson field  
\begin{equation}
S_B(\phi, \phi^*)=\int^{\beta}_0 d\tau\int d{\bf x}\Big[\phi^*({\bf x},\tau)
\Big(\frac{\partial}{\partial \tau}-\frac{\nabla^2}{2m_B}-\mu_B\Big)
\phi({\bf x},\tau)+\frac{1}{2}\gamma|\phi({\bf x},\tau)|^4\Big]\, ,
\label{sbos}
\end{equation} 
the action for the fermion field  
\begin{equation}
S_F(\psi, \psi^*)=\int^{\beta}_0 d\tau\int d{\bf x}\Big[\psi^*({\bf x},\tau)
\Big(\frac{\partial}{\partial \tau}-\frac{\nabla^2}{2m_F}-\mu_F\Big )
\psi({\bf x},\tau)\Big]
\label{sfermi}
\end{equation} 
and the interaction term containing both fields     
\begin{equation}
S_I(\phi, \phi^*,\psi,\psi^*)=\lambda\int^{\beta}_0 d\tau\int d{\bf x}\Big[
\psi^*({\bf x},\tau)\psi({\bf x},\tau)\phi^*({\bf x},\tau)
\phi({\bf x},\tau)\Big]\, .
\label{sinter}
\end{equation}
Here $\mu_B$ and $\mu_F$ denote the boson and fermion chemical potentials 
respectively. In writing the action functional, the usual zero--range 
approximation for the interactions between atoms has been adopted 
\cite{Bij00,Hei00,Efr02,Viv02}. The corresponding coupling constants 
are defined as follows: 
$\gamma=4\pi a_{BB}/m_B$ and $\lambda=2\pi a_{BF}/m_R$, 
where $a_{BB}$ and $a_{BF}$ are the boson--boson and fermion--boson 
scattering lengths respectively, and $m_R=m_Bm_F/(m_B+m_F)$ 
is the reduced mass for a boson of mass $m_B$ and a fermion of mass $m_F$. 
Finally, according to the previous remarks, in Eq. (\ref{sfermi})  
the fermion--fermion interaction has been omitted. 
\par
The Bose gas is treated within the Gaussian approximation, 
i.e. all terms of order higher than second in the fluctuations 
of the boson field about the uniform and constant condensate 
amplitude are neglected. This corresponds to the Bogoliubov 
approximation for a simple Bose gas. For a weakly interacting gas,   
$n_B\,a_{BB}^3<<1$ (~$n_B$~is the boson density~) and at low temperatures,  
this approximation is well justified \cite{Walecka}. The first 
step consists in evaluating the condensate amplitude about which 
the perturbative expansion can be performed. The standard  
methods employed for a simple Bose gas \cite{Negele,Sto99} 
are extended to a fermion--boson mixture. First 
the action functional is integrated over the fermion fields,  
then, a uniform and constant boson field $\phi({\bf x},\tau)=\varphi$,  
which makes the action stationary, is searched for. 
The integration over the fermion fields can be performed explicitely, 
since these fields occur with a Gaussian form in the functional integral. 
This gives the two  following contributions to the action functional 
\begin{equation}
-{\rm Tr}\,\ln {G}_0^{-1}({\bf x-x}^\prime,\tau-\tau^\prime)
-{\rm Tr}\,\ln\Big({\bf 1}+\lambda
{G}_0^{-1}({\bf x-x}^\prime,\tau-\tau^\prime) 
\phi({\bf x},\tau)\phi^*({\bf x},\tau)\Big)\, ,
\label{fermint}
\end{equation}
where the trace operation implies integration over the continuous variables 
${\bf x},\tau$. Here ${G}_0^{-1}({\bf x-x}^\prime,\tau-\tau^\prime)$ 
is the inverse of the Green function for free fermions. The first variation 
of the action functional takes the contribution 
\[
-\lambda\,\phi({\bf x},\tau)\,{G}({\bf x},{\bf x},\tau,\tau+0^+)=
\lambda\,\phi({\bf x},\tau)n_F ,\] 
from the second term of Eq. (\ref{fermint}). The quantity 
${G}({\bf x},{\bf x}^\prime,\tau,\tau^\prime)$ is the 
Green function for fermions interacting with the scalar field 
$\phi({\bf x},\tau)$. Adding the contributions coming from the boson term 
of the action, for the density of the zero--momentum Bose 
condensate, $n^0_B=\varphi^2$, the following equation is obtained  
\begin{equation}
n^0_B=\,\frac{\mu_B-\lambda n_F}{\gamma}\,.
\label{condens}
\end{equation}
This equation has been derived assuming, without loss of generality, 
that the quantity $\varphi$ is real. We can see that the effect 
of the interaction with the Fermi gas simply amounts 
to replacing the chemical potential $\mu_B$ with the effective 
value $\mu_B-\lambda n_F$. \par 
The next step is the introduction of the change of variables 
$\phi^\prime=\phi-\varphi$ in the terms  $S_B(\phi,\phi^*)$ 
and $S_I(\phi, \phi^*,\psi,\psi^*)$ of the action functional. 
The term with $\varphi^2$ represents the interaction of the fermions 
with the condensate mean--field, $\lambda n_B^0$ and can be added to   
the chemical potential $\mu_F$ in $S_F(\psi, \psi^*)$.  
The terms containing products of more than two amplitudes 
$\phi^\prime$, are neglected according to the Bogoliubov 
approximation. Thus, the action for the fluctuations of the 
boson field is given by 
\begin{eqnarray}
S^{(2)}(\phi^\prime,\phi^{\prime *},\psi,\psi^*)
=&& \frac{1}{2}\int^{\beta}_0
d\tau\,d\tau^\prime\int d{\bf x}d{\bf x}^\prime\,{\hat \phi}^{\prime \dag}
({\bf x},\tau)  
{\widehat D}^{-1}({\bf x},{\bf x}^\prime,\tau,\tau^\prime)
{\hat \phi}^{\prime}({\bf x}^\prime,\tau^\prime)
\nonumber
\\
&&+ \frac{1}{2}\int^{\beta}_0d\tau\int d{\bf x}
\Big({\hat \phi}^{\prime \dag}({\bf x},\tau) 
{\hat \Gamma}({\bf x},\tau)+c.c.\Big)\, ,
\label{bogo}
\end{eqnarray}
where the vector field 
\[{\hat \phi}^{\prime}({\bf x},\tau)= \left( \begin{array} {c} 
                            \phi^\prime({\bf x},\tau)\\
                            \phi^{\prime *}({\bf x},\tau)
                            \end{array} \right)\,,\]       
the source term
\[{\hat \Gamma}({\bf x},\tau)= \lambda\varphi 
                            \left( \begin{array} {c} 
                       \psi^*({\bf x},\tau)\psi({\bf x},\tau)\\
                        \psi^*({\bf x},\tau)\psi({\bf x},\tau)
                            \end{array} \right)\, ,\]       
and the inverse of the Green function for the fluctuations 
$\phi^\prime$  
\[{\widehat D}^{-1}({\bf x},{\bf x}^\prime,\tau,\tau^\prime)
={\widehat D}^{-1}_0({\bf x},{\bf x}^\prime,\tau,\tau^\prime)+
{\widehat V}({\bf x},{\bf x}^\prime,\tau,\tau^\prime)\, \]
are introduced with the inverse of the Bogoliubov propagator 
\[{\widehat D}^{-1}_0({\bf x},{\bf x}^\prime,\tau,\tau^\prime)=
\left( \begin{array} {cc} 
{\displaystyle \frac{\partial}{\partial \tau}-\frac{\nabla^2}{2m_B}}
          +\gamma n^0_B &\gamma n^0_B  \\
          \gamma n^0_B & {\displaystyle -\frac{\partial}{\partial \tau}
          -\frac{\nabla^2}{2m_B}}+\gamma n^0_B 
                            \end{array} \right)\delta(\tau-\tau^\prime)
                           \delta({\bf x}-{\bf x}^\prime)\]
and the coupling matrix 
\[{\widehat V}({\bf x},{\bf x}^\prime,\tau,\tau^\prime)=\lambda
\left( \begin{array} {cc}
               \psi^*({\bf x},\tau)\psi({\bf x},\tau) & 0 \\
               0 & \psi^*({\bf x},\tau)\psi({\bf x},\tau) 
                         \end{array} \right)\delta(\tau-\tau^\prime)
                          \delta({\bf x}-{\bf x}^\prime)  \, .\]
In order to obtain an effective interaction for the fermions, the integration  
over the fluctuations  $\phi^{\prime}$ and $\phi^{\prime *}$ should be 
performed. The result is an effective action containing only  
the fermionic fields $\psi$ and $\psi^*$ 
\begin{eqnarray}
S_{eff}(\psi,\psi^*)=&&\int^{\beta}_0 d\tau
\int d{\bf x}\Big[\psi^*({\bf x},\tau)
\Big(\frac{\partial}{\partial \tau}-\frac{\nabla^2}{2m_F}
-(\mu_F-\lambda n_B^0)\Big )
\psi({\bf x},\tau)\Big]
\nonumber
\\
&&+\frac{1}{2}
\ln({\rm det}{\widehat D}_0^{-1})
+\frac{1}{2}
\ln\big[{\rm det}({\bf 1}+{\widehat D}_0{\widehat V})\big]
\nonumber
\\
&&-\frac{1}{2}
\int^{\beta}_0d\tau d\tau^\prime\int d{\bf x}d{\bf x}^\prime\,
{\hat \Gamma}^\dag({\bf x},\tau)
{\widehat D}({\bf x},{\bf x}^\prime,\tau,\tau^\prime)
{\hat \Gamma}({\bf x}^\prime,\tau^\prime)
\, .
\label{effact1}
\end{eqnarray}
The term $\ln({\rm det}{\widehat D}_0^{-1})$ together with the 
action evaluated at the stationary point $\phi^{\prime}=0$, 
give the grand potential in the Bogoliubov approximation   
for a Bose gas with the effective chemical potential $\mu_B-\lambda n_F$. 
\par
The last two terms on right hand side 
of Eq. (\ref{effact1}) represent an effective 
interaction among fermions. When expanded in powers 
of the fermion fields, both terms give rise to $n$--body forces. 
Only terms containing up to four fields are taken into account, 
so both states with more than two virtual phonons in the 
boson--density fluctuations and $n$--body forces among fermions, 
with $n$ greater than two, are neglected. We expect this to be a reasonable 
approximation for sufficiently dilute gases. According to this 
approximation, the boson Green function is replaced with the Bogoliubov 
propagator, ${\widehat D}={\widehat D}_0$ in the source term,     
and terms up to second order are retained in the expansion 
\begin{equation}
\frac{1}{2}{\rm Tr}\ln({\bf 1}+{\widehat D}_0{\widehat V})=
\,-\frac{1}{2}\sum_n\frac{(-1)^n}{n}{\rm Tr}
\big({\widehat D}_0{\widehat V}\big)^n\,.
\label{log}
\end{equation}
\par
The space--Fourier transform of the contribution from the 
source term of Eq. (\ref{effact1}) to the effective interaction 
between fermions is given by  
\begin{equation}
V_1(q,\tau_1-\tau_2)=-\lambda^2n^0_B\big[D_0^{(11)}(q,\tau_1-
\tau_2)+D_0^{(11)}(q,\tau_2-\tau_1)+2D_0^{(12)}(q,\tau_1-\tau_2)\big]\, ,
\label{v1}
\end{equation}
where $D_0^{(ij)}$ are the elements of the $2\times2$ matrix 
${\widehat D}_0$; for their explicit expressions see, e.g., 
Ref. \cite{Walecka} (~the definition of ${\widehat D}_0$ used here 
differs from that of Ref. \cite{Walecka} by an overall minus sign~).  
Concerning the contributions from the expansion of Eq. (\ref{log}), 
the term with $n=1$ represents the mean field acting on the fermions 
due to the noncondensate fraction of the Bose gas. This term 
can be added to the chemical potential of the fermions, giving together 
with the analogous term of Eq. (\ref{effact1}) the effective chemical 
potential $\mu^*_F=\mu_F-\lambda n_B$ for the fermions. The space--Fourier 
transform of the contribution from the term with $n=2$ reads 
\begin{eqnarray}
V_2(q,\tau_1-\tau_2)=-\lambda^2\int\frac{d{\bf k}}{(2\pi)^3}
&&\big[D_0^{(11)}(k,\tau_1-\tau_2)D_0^{(11)}({\bf k}-{\bf q},\tau_2-\tau_1)
\nonumber
\\
&&+
D_0^{(12)}(k,\tau_1-\tau_2)D_0^{(12)}({\bf k}-{\bf q},\tau_2-\tau_1)\big]\,. 
\label{v2}
\end{eqnarray}
This term is the contribution to the effective interaction 
from states containing two virtual phonons. 
\par
A rough estimate of the terms of Eq. (\ref{log}) corresponding to  
the exchange of $n>2$ virtual phonons between fermions  
can be done by means of the following considerations. 
The expansion of Eq. (\ref{log}) contains correlations of 
the boson density fluctuations of higher and higher order when $n$ 
increases. According to the basic assumption made in this paper, the 
correlations will be evaluated at equal times. In this case, we should 
expect that the contribution of a term of the series with 
respect to that of the previous term, contains an extra power of 
$\lambda n^{\prime}_B$. Here, $n^{\prime}_B$ is the density of the 
noncondensate bosons. In addition, when $n$ increases by a unit, the 
effective action of Eq. (\ref{effact1}) contains an additional 
integration on the time, which gives a multiplicative factor of the 
order of magnitude as $\beta$. Thus, we can interpret the right hand 
side of Eq. (\ref{log}) as a perturbative expansion in powers of the 
adimensional parameter $\lambda n^{\prime}_B/T$. Similar considerations 
can be done for the terms neglected in the perturbative expansion of 
the last term of Eq. (\ref{log}). At the densities and temperatures 
considered in this paper the parameter $|\lambda n^{\prime}_B/T|$ 
takes a value of $\sim 0.15$. Since the order parameter of the 
superconducting phase, in the approach of this paper turns out to be not 
critically dependent on the fermion--fermion interaction (~see 
Eq. (\ref{pairl1}) of Sec. III~), corrections to the effective 
fermion--fermion interaction coming from terms of 
order higher than second will be neglected in order to simplify 
calculations. 
\subsection{Pairing field}
The effective interaction obtained involves fermions at two different 
times. The approximations which can be introduced in order to 
get rid of the retardation effects and simplify calculations, 
will be discussed later. Now, the equations for the effective field 
representing the pairing of two fermions, are derived. 
\par
With the approximations discussed above, the partition function 
of Eq. (\ref{grpart}) assumes the following factorized form 
\begin{equation} 
Z={\rm exp}(-\beta \Omega_B)\int\,{\cal D}(\psi,\psi^*)\,{\rm exp}
\big[-S_{eff}(\psi,\psi^*)\big]\,,
\label{zfact}
\end{equation}
where $\Omega_B$ is the grand potential of the Bose gas, while 
the fermion effective action of Eq. (\ref{effact1}) becomes 
\begin{eqnarray}
S_{eff}(\psi,\psi^*)=&&\int^{\beta}_0 d\tau
\int d{\bf k}\Big[\psi^*({\bf k},\tau)
\Big(\frac{\partial}{\partial \tau}+\frac{k^2}{2m_F}
-\mu^*_F\Big)
\psi({\bf k},\tau)\Big]
\nonumber
\\
&&+
\frac{1}{2(2\pi)^3}\int_0^\beta d\tau_1d\tau_2\int\prod_i\,d{\bf k}_i
\,\delta({\bf k}_1+{\bf k}_2-{\bf k}_3-{\bf k}_4)
\nonumber
\\
&&\times
\Big[V_{eff}({\bf k}_3-{\bf k}_1,\tau_1-\tau_2)
\psi^*({\bf k}_1,\tau_1)\psi^*({\bf k}_2,\tau_2)
\psi({\bf k}_4,\tau_2)\psi({\bf k}_3,\tau_1)\Big]
\, ,
\label{effact2}
\end{eqnarray}
with $V_{eff}({\bf k}_3-{\bf k}_1,\tau_1-\tau_2)=
V_1(|{\bf k}_3-{\bf k}_1|,\tau_1-\tau_2)+
V_2(|{\bf k}_3-{\bf k}_1|,\tau_1-\tau_2)$. 
\par
The pairing field is identified with the composite field which 
makes the effective action $S_{eff}(\psi,\psi^*)$  stationary. 
This can be determined by means of a Stratonovich--Hubbard  
transformation \cite{Stra58,Hub59} introducing auxiliary fields for 
the couples $\psi\psi$ and $\psi^*\psi^*$, and by equating to   
zero the first variation of the resulting effective action with  
respect to the auxiliary fields. After the transformation,  
the exponential in the integral of Eq. (\ref{zfact}) is replaced by  
\begin{eqnarray}
{\rm exp}\big[-S_{eff}(\psi,\psi^*)\big]=&&{\rm exp}
\bigg[-\int^{\beta}_0 d\tau
\int d{\bf k}\psi^*({\bf k},\tau)
\Big(\frac{\partial}{\partial \tau}+\frac{k^2}{2m_F}
-\mu^*_F\Big)
\psi({\bf k},\tau)\bigg]
\nonumber
\\
&&\times \int{\cal D}(\Delta^*,\Delta)
{\rm exp}\bigg[\int^{\beta}_0\prod_i\,d\tau_i\int\prod_i\,d{\bf k}_i
\nonumber
\\
&&\frac{1}{2}\Big[\Delta^*({\bf k}_1,{\bf k}_2,\tau_1,\tau_2)V^{-1}_{eff}
({\bf k}_1,{\bf k}_2,\tau_1,\tau_2;{\bf k}_3,{\bf k}_4,\tau_3,\tau_4)
\Delta({\bf k}_3,{\bf k}_4,\tau_3,\tau_4)\Big]
\nonumber
\\
&&+\frac{1}{2}\int d{\bf k}_1d{\bf k}_2\int d\tau_1d\tau_2\Big[
\Delta^*({\bf k}_1,{\bf k}_2,\tau_1,\tau_2)\psi({\bf k}_1,\tau_1)
\psi({\bf k}_2,\tau_2)
\nonumber
\\
&&+\psi^*({\bf k}_1,\tau_1)
\psi^*({\bf k}_2,\tau_2)\Delta({\bf k}_1,{\bf k}_2,\tau_1,\tau_2)\Big]
\bigg]\, ,
\label{aux1}
\end{eqnarray}
where $\Delta$ and $\Delta^*$ are bosonic auxiliary fields and the matrix  
\[V^{-1}_{eff}({\bf k}_1,{\bf k}_2,\tau_1,\tau_2;
{\bf k}_3,{\bf k}_4,\tau_3,\tau_4)\] 
is the inverse of the effective interaction 
\[<{\bf k}_1{\bf k}_2|V_{eff}(\tau_1,\tau_2;\tau_3,\tau_4)|{\bf k}_3{\bf k}_4>
=\]
\[\frac{1}{(2\pi)^3}V_{eff}({\bf k}_3
-{\bf k}_1,\tau_1-\tau_2)\delta(\tau_3-\tau_1)
\delta(\tau_4-\tau_2)\delta({\bf k}_1+{\bf k}_2-{\bf k}_3-{\bf k}_4)\, .\] 
In the functionals containing the coupling between fermions and 
auxiliary fields, only the parts of $\Delta$ and $\Delta^*$ 
which are antisymmetric under the permutation $1\leftrightarrow 2$ 
survive. This reflects the prescription of the Pauli principle 
for fermions in the same spin state. 
\par
The integration over the fermionic fields can be formally performed, 
giving the following action for the auxiliary fields 
\begin{eqnarray}
S(\Delta,\Delta^*)=-\int^{\beta}_0\prod_i\,d\tau_i\int\prod_i\,d{\bf k}_i
\frac{1}{2}
\Big[&&\Delta^*({\bf k}_1,{\bf k}_2,\tau_1,\tau_2)V^{-1}_{eff}
({\bf k}_1,{\bf k}_2,\tau_1,\tau_2;{\bf k}_3,{\bf k}_4,\tau_3,\tau_4)
\nonumber
\\
&&\times\Delta({\bf k}_3,{\bf k}_4,\tau_3,\tau_4)\Big]
-\frac{1}{2}{\rm Tr\,ln}({\bf 1}-{\widehat{\cal G}}_0{\widehat \Delta})
\, ,
\label{aux2}
\end{eqnarray}
where ${\widehat{\cal G}}_0$ is a $2\times 2$ diagonal matrix, whose 
elements are the propagator for independent fermions with 
the effective chemical potential $\mu^*_F$, and the matrix 
${\widehat \Delta}$ is given by 
\[{\widehat \Delta}({\bf k}_1,{\bf k}_2,\tau_1,\tau_2)
=\left( \begin{array} {cc}
               0 & \Delta({\bf k}_1,{\bf k}_2,\tau_1,\tau_2)\\
               \Delta^*({\bf k}_1,{\bf k}_2,\tau_1,\tau_2) & 0 
                         \end{array}\right) \, .\]

The equation for the pairing field can be obtained by equating to 
zero the functional derivative of $S(\Delta,\Delta^*)$ with respect 
to either $\Delta^*$ or $\Delta$, equivalently. It reads 
\begin{eqnarray}
\Delta({\bf k},{\bf k}^\prime,\tau-\tau^\prime)=\frac{1}{(2\pi)^3}
\int d{\bf k}_1d{\bf k}_2&&V_{eff}({\bf k}_1-{\bf k},\tau-\tau^\prime)
\delta({\bf k}+{\bf k}^\prime-{\bf k}_1-{\bf k}_2)
\nonumber
\\
&&\times{\cal G}^{(12)}({\bf k}_2,{\bf k}_1,\tau-\tau^\prime)\,,
\label{pair1}
\end{eqnarray}
where ${\cal G}^{(12)}({\bf k}_2,{\bf k}_1,\tau-\tau^\prime)$ 
is the anomalous propagator for fermions interacting with the pairing 
field ${\widehat \Delta}$ (~see, e.g., Ref. \cite{Walecka}~). 
Because of time--translation invariance  
both the pairing field and the fermion propagator depend only on 
the difference of times. 
\par
It is convenient to introduce the center of mass momentum 
${\bf P}={\bf k}+{\bf k}^\prime={\bf k}_1+{\bf k}_2$, thus, 
for a given value of ${\bf P}$, an equation containing only one 
integration variable, ${\bf k}$, can be obtained. 
The solutions of Eq. (\ref{pair1}) with ${\bf P}\not=0$, 
correspond to a breaking of the space--translational symmetry, 
i.e. the LOFF phase \cite{Loff}. Here, only solutions with vanishing 
center of mass momentum are considered. 
In this case, the pairing field and the anomalous propagator 
depend on only one momentum, ${\bf k}=-{\bf k}^\prime$ and  
${\bf k}_1=-{\bf k}_2$, respectively. The propagator 
${\cal G}^{(12)}({\bf k}_1,\tau-\tau^\prime)$ 
is given by the Fourier series 
\begin{equation}
{\cal G}^{(12)}({\bf k}_1,\tau-\tau^\prime)=-\frac{1}{\beta}
\sum_n\frac{e^{-i\omega_n(\tau-\tau^\prime)}\Delta({\bf k}_1,\omega_n)}
{\omega_n^2+\xi^{2}_{k_1}
+\Delta({\bf k}_1,\omega_n)\Delta^*({\bf k}_1,\omega_n)}
\,,
\label{g12}
\end{equation}
where $\omega_n$ are the Matsubara frequencies for fermions and 
$\xi^{2}_{k_1}=(k^2_1/2m_F-\mu^*_F)^2$. 
\par
Now, two different paths of approximation which allow us to neglect the 
time (\,or frequency\,) dependence of the pairing field, are discussed. 
Generally, for evaluating the time behavior of a propagator we  
should make an analytical continuation to real times 
(\,$-i\tau\rightarrow t$\,) or, equivalently to   
real frequencies (\, $i\omega_n\rightarrow \omega$\,). 
However, for the two cases that will be envisaged, we can directly 
refer to the equations just derived. Both approximations are  
based on the comparison of the characteristic times of propagation 
for fermions and for bosons. For a quasi degenerate Fermi gas only the 
particles with momenta lying about the Fermi surface partecipate in 
the pairing process. Then, for a given momentum $q$, the fermion  
time--scale can be estimated as $t_F=1/v_Fq$. While for bosons,  
the quantity $t_B=1/c_sq$, where $c_s$ is the sound velocity in the 
Bose gas, can be assumed to give an estimate of the characteristic time. 
The comparison between the two time--scales reduces to compare 
the velocities $v_F$ and $c_s$. The case, when $c_s>>v_F$, is 
examined first. This is the assumption usually made when studying 
boson--induced interactions between fermions for Bose--Fermi 
mixtures. For $c_s>>v_F$ retardation 
effects on the induced interaction can be neglected, i.e. only 
the $\omega=0$ component of the time--Fourier transform of 
the boson propagators in Eqs. (\ref{v1}) and (\ref{v2}) is 
retained. Also, the fermion propagator 
${\cal G}^{(12)}({\bf k},\tau-\tau^\prime)$ can be evaluated at equal 
times (\,$\tau=\tau^\prime$\,). The pairing field in turn becomes 
an instantaneous function of the time difference, 
$\Delta({\bf k},\tau-\tau^\prime)=\Delta({\bf k})\delta 
(\tau-\tau^\prime)$. The equation obeyed by the pairing field is the usual 
gap equation of the BCS theory. The second approximation is the opposite 
of that just discussed: the Fermi velocity can be assumed to be 
much larger than the sound velocity. Then, the propagator 
${\cal G}^{(12)}({\bf k},\tau-\tau^\prime)$ is a fast changing  
function of time with respect to the boson propagators, so that 
in Eq. (\ref{pair1}) the static limit of the effective interaction 
can be taken. In this case, the boson--induced interaction is 
the space--Fourier transform ( times $-\lambda^2$ ) of the static 
correlation function for density fluctuations. In the following 
calculations, this second approximation will be adopted, since it is  
closer to the  physical situations realized in   
experiments. Thus, in the frequency representation, the equation 
(\ref{pair1}) for the pairing field with ${\bf P}=0$,  
is given by the following expression  
\begin{equation}
\Delta({\bf k},\omega_n)=-\int\frac{d{\bf k}_1}{(2\pi)^3}
\frac{V_{eff}({\bf k}-{\bf k}_1,\tau=\tau^\prime)
\Delta({\bf k}_1,\omega_n)}
{\omega_n^2+\xi_{k_1}^2+|\Delta({\bf k}_1,\omega_n)|^2}\, .
\label{pair2}
\end{equation}
When evaluating the energy gap in the spectrum of 
elementary excitations, an analytical 
continuation of this equation to real frequencies 
should be made, in order to determine the real poles 
of the fermion Green function. However, for 
assessing the occurrence of a superconducting phase and for evaluating 
the order of magnitude of the energy gap, it is sufficient to take 
into account only the $\omega=0$ component of the pairing field,  
$\Delta({\bf k})$. This quantity satisfies the equation 
\begin{equation}
\Delta({\bf k})=-\int\frac{d{\bf k}_1}{(2\pi)^3}
\frac{V_{eff}({\bf k}-{\bf k}_1)\Delta({\bf k}_1)}
{\xi_{k_1}^2+|\Delta({\bf k}_1)|^2}\, .
\label{pair}
\end{equation}
Here $V_{eff}({\bf k}-{\bf k}_1)$ denotes the effective 
interaction at equal times, its explicit expression, 
with $q=|{\bf k}-{\bf k}_1|$, is  
\begin{eqnarray}V_{eff}(q)=&&-\lambda^2 n^0_B\Big[1+2n_B^\prime(q)
-\frac{\gamma n^0_B}{E_q}\big[1+2{\tilde n}(q)\big]\Big]
-\lambda^2\,\int\frac{d{\bf k}}{(2\pi)^3}\Big[n_B^\prime(k)
\big[1+n_B^\prime(|{\bf k}-{\bf q}|)\big]+
\nonumber
\\
&&+\frac{(\gamma n^0_B)^2}
{4E_kE_{|{\bf k}-{\bf q}|}}\big[1+2{\tilde n}(k)\big]
\big[1+2{\tilde n}(|{\bf k}-{\bf q}|)\big]\Big]\, ,
\label{veff}
\end{eqnarray}
where $E_k$ are the excitation energies of the Bose gas 
calculated within the Bogoliubov approximation, 
$E_k=\sqrt{\big(k^2/(2m_B)\big)^2+2\gamma n_B^0\,k^2/(2m_B)}$, 
$n_B^\prime(k)$ the occupation numbers of noncondensate 
bosons, and 
\[{\tilde n}(k)=\frac{1}{e^{\beta E_k}-1}\,.\]
The first term on right hand side of Eq. (\ref{veff}) 
is the contribution from states containing one virtual phonon, and 
it comes from the correlations of the products of the condendate and 
noncondensate amplitudes taken at two different space points. The 
second term is the contribution from states containing two 
virtual phonons, and it is related to the noncondensate--density 
correlations. 
\par
A few remarks should be made. First, according to the Pauli 
principle, the pairing field 
should be an odd function of ${\bf k}$ in order to change sign  
under the permutation of the two correlated fermions. This excludes 
contributions to the pairing field from interactions acting only 
in the $s$--wave channel. So that, as previously remarked, 
the bare fermion--fermion interaction can be completely neglected 
in the present scheme. In addition, the function $V_{eff}(q)$ 
has a simple pole, when $q\rightarrow 0$. 
However, this is a fictitious pole, since, if the 
the integrals in Eq. (\ref{veff}) are replaced by a sum over the momenta, 
the boson states with vanishing momentum should be excluded. Actually, 
this restriction may be omitted, because it affects the integrand of 
Eq. (\ref{pair}) only at a single point and the integration gives a finite 
result. Finally, it is worth noting that $V_{eff}(q)$ 
does not represent a usual two--body potential, since an integration 
over the time is absent. 
\par
In order to solve Eq. (\ref{pair}) it is convenient to 
expand $\Delta({\bf k})$ in spherical harmonics 
\[\Delta({\bf k})=\sum_{l,m}\sqrt{\frac{4\pi}{2l+1}}
\Delta_{l,m}(k)Y_l^m(\Omega_k)\,,\] 
with only odd values of $l$ contributing. 
To simplify calculations, an angle--average 
approximation for  $|\Delta({\bf k}_1)|^2$ can be made  
\begin{equation}
|\Delta({\bf k}_1)|^2\rightarrow \bar\Delta^2(k_1)=
\sum_{l,m}\frac{1}{2l+1}|\Delta_{l,m}(k_1)|^2\,.
\label{deltaeff}
\end{equation} 
With this approximation, the different $m$--components 
turn out to be uncoupled and all equal; in addition, 
they can be assumed to be real and positive, without loss of generality.   
The quantity $\bar\Delta^2(k)$ represents the averaged energy gap 
in the quasiparticle spectrum of the superconducting phase. \par
The equations for the partial waves of the pairing field,  
\begin{equation}
\Delta_l(k)=-\int\frac{dk_1}{4\pi^2}k_1^2\frac{V_l(k,k_1)\Delta_l(k_1)}
{\xi_{k_1}^2+\bar\Delta^2(k_1)}\,,
\label{pairl}
\end{equation}
can still be coupled via the term $\bar\Delta^2(k_1)$. Here, 
the components $V_l(k,k_1)$ are defined as   
\[V_l(k,k_1)=\int dx\,V_{eff}(k^2+k_1^2-2kk_1x)P_l(x)\,.\]
They are bounded quantities for $k=k_1$, in spite of the occurence 
of a pole when the argument of $V_{eff}(q)$ vanishes. 
\section{Results}
The temperature, density and relative concentrations of  
the gas mixture are assumed such that a relevant fraction of 
the Bose gas is in the condensate phase and the Fermi gas is still  
quite degenerate. To be more specific, we refer to the experimental 
environment described in Ref. \cite{Roa02}, where a mixture of  
$^{87}{\rm Rb}$ (\,bosons\,) and $^{40}{\rm K}$ (\,fermions\,) 
atoms, with a relevant ratio between 
the fermion and boson densities, $n_F/n_B\sim 0.5$, 
has been studied. For the homogeneous boson gas, 
a value $T_c=110 {\rm n\,K}$ for the critical temperature of 
condensation has been chosen, in order to reproduce values of 
temperature and condensate fraction close to those reported in 
Ref. \cite{Roa02}. The corresponding boson density is 
$n_B=1.45\times10^{13}{\rm cm}^{-3}$. This is a reasonable value 
for the average boson density obtained in the experiment of 
Ref. \cite{Roa02}. Moreover, for the boson--boson and 
boson--fermion scattering lengths, the values $a_{BB}=113 a_0$ 
and $a_{BF}=-330 a_0$ are used, where $a_0$ is the 
Bohr radius. The range of the condensate fractions considered in the 
present calculations is $0.4\le n_B^0/n_B\le 0.8$, and the corresponding 
interval of temperatures is $46{\rm n\,K}\le T\le 87{\rm n\,K}$.  
With these values of the relevant parameters, the ratio 
between the Fermi velocity, defined as $v_F=\sqrt{2\,\mu^*_F/m_F}$, 
and the sound velocity in the Bose gas, $c_s=\sqrt{\gamma n_B^0/m_B}$, 
turns out to be about $20$. This value of $v_F/c_s$ is 
sufficiently large to justify the approximation of Eq. (\ref{veff}) 
to the effective interaction. Moreover, the Fermi temperature 
corresponding to the density $n_F\sim 0.5n_B$, $T_F\sim 320{\rm n\,K}$,   
is high enough with respect to the temperatures in the interval 
considered, so that the Pauli principle is still operating. 
\par
The results for the effective interaction are discussed first. 
A simple inspection of Eq. (\ref{veff}) shows that the 
contributions from states with one virtual phonon and from 
states with two virtual phonons, approach the values   
$-\lambda^2\,n_B^0$ and $-\lambda^2\,n_B^\prime$, 
respectively, when $q\rightarrow\infty$.  
\par
\begin{figure}
\includegraphics{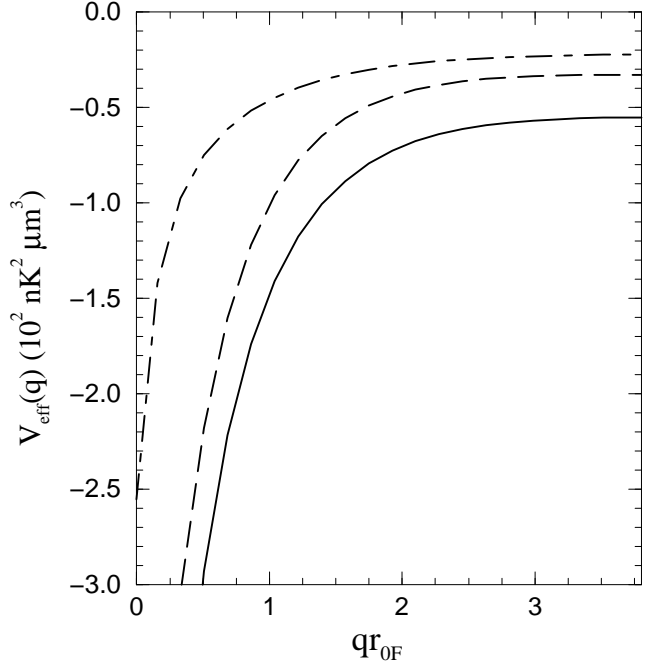}
\caption{\label{fig1}Induced effective interaction between fermions as a 
function of the momentum transfer $q$, for a mixture of $^{40}{\rm K}$ 
and $^{87}{\rm Rb}$ (solid line), $r_{0F}$ is the mean spacing between 
fermions. The contributions from states 
with one virtual phonon (~dashed line~) and with two virtual phonons   
(~dot--dashed line~) are also shown. The condensate fraction is 
$n_B^0/n_B=0.6$.}
\end{figure}

In Fig.~\ref{fig1} the effective interaction of Eq. (\ref{veff}) is shown 
togheter with the separate contributions from one--phonon   
and two--phonon states, for $n_B^0/n_B=0.6$. We can see that 
the two contributions are of the same order of magnitude 
for $q\geq 0.5\,r_{0F}^{-1}$, with $r_{0F}$ denoting the mean 
spacing of the Fermi atoms, and that the asymptotic values are practically   
reached already at $q\sim 1.5\,r_{0F}^{-1}$. This  suggests that the 
contribution from states containing two virtual phonons cannot be 
neglected, unless the condensate depletion is very small. 
\par
\begin{figure}
\includegraphics{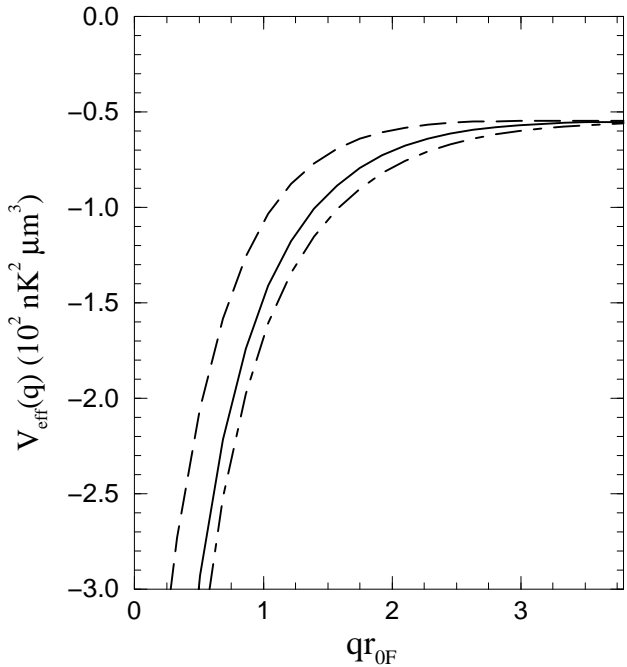}
\caption{\label{fig2}Induced effective interaction between 
fermions as a function of the momentum transfer $q$, for three 
different values of the condensate fraction. The Bose--Fermi mixture 
is the same as in Fig.~\ref{fig1}.  
Dashed line: $n_B^0/n_B=0.8$; solid line: 
$n_B^0/n_B=0.6$; dot--dashed line: $n_B^0/n_B=0.4$.}
\end{figure}

In Fig.~\ref{fig2} the effective interaction is displayed for three 
different values of the condensate fraction. From Fig.~\ref{fig2} 
we can see that, in the considered range of $n_B^0/n_B$, 
for a fixed value of the boson density the magnitude of 
the effective interaction increases with the condensate depletion, 
or, equivalently, with the temperature. This is due to 
the fact that the increase with the noncondensate fraction 
of the two--phonon contribution, overcomes the decrease with $n_B^0$ 
of the one--phonon contribution. However, for higher values of the condensate  
depletion, the increase of $V_{eff}(q)$ results considerably quenched. 
\par
Concerning the solutions of the gap equation (\ref{pairl}),  
it should first be remarked that the 
fast approach of the effective interaction to an 
almost constant value makes the odd components $V_l(k,k_1)$ 
vanish rapidly at high momenta. As a consequence, for evaluating 
the integral of Eq. (\ref{pairl}) it is not necessary 
to introduce any cut--off on high values of $k_1$. The explicit 
solutions of Eq. (\ref{pairl}) display two general features for the 
physical situations considered: they give, for all values of $l$, 
a small energy gap, with respect to the effective chemical potential of the  
of the fermions, $\Delta_l(k)/\mu^*_F\sim 10^{-2}$, and they give 
a non vanishing energy gap only for a single value of $l$ at a time, 
without any coupling to the remaining components. The 
latter feature can be better understood by observing that, for small values  
of $\Delta_l(k)$, the most relevant contribution to the integral of 
Eq. (\ref{pairl}) comes from a small domain about the effective 
Fermi momentum, $k^*_F=\sqrt{2\mu^*_F\,m_F}$. This allows us to 
approximate all the quantities in the integrand, apart from $\xi_{k_1}$, 
by the values that they take at $k_1=k^*_F$. Thus, Eq. (\ref{pairl}) 
turns into the following closed equation for each value of $k$ 
\begin{equation}
\Delta_l(k)=-\frac{1}{4\pi^2}m_Fk^*_F\,V_l(k,k^*_F)\Delta_l(k^*_F)\frac{1}
{\sqrt{\bar\Delta^2(k^*_F)}}\,\Big[\frac{\pi}{2}+\arctan(\frac{\mu^*_F}
{\sqrt{\bar\Delta^2(k^*_F)}})\Big]\,,
\label{pairl1}
\end{equation}
where the angle average $\bar\Delta^2(k^*_F)$, (\,see Eq. (\ref{deltaeff})\,), 
is given by 
\[\bar\Delta^2(k^*_F)=\sum_{l^\prime}\Delta_{l^\prime}^2(k^*_F)\,.\]
It is easy to see that Eq. (\ref{pairl1}) can have nontrivial solutions 
only when $\Delta_{l^\prime}=0$ for $l^\prime\not=l$. In addition, 
it is worth noting that the energy gaps given by Eq. (\ref{pairl1}) 
represent a very satisfactory approximation to the numerical solutions of 
Eq. (\ref{pairl}). 
\par
The solutions of Eq. (\ref{pairl}) indicate that transitions 
to different superconducting phases could occur. Each phase is 
characterized by a nonisotropic pairing field, whose angular  
dependence is given by a specific spherical harmonic. 
The corresponding energy gap contains only one term $\Delta_l(k)$. 
\par
\begin{figure}
\includegraphics{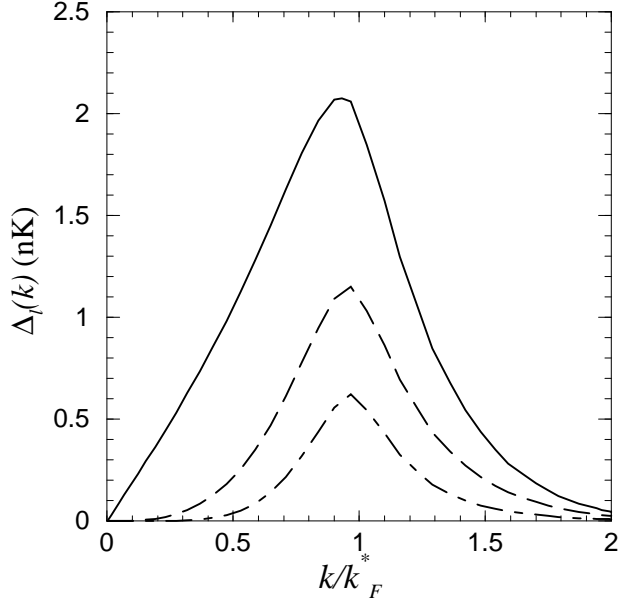}
\caption{\label{fig3}Energy gap for different $l$--channels,  
as a function of the wave vector $k$ measured in units of 
$k^*_F=\sqrt{2\mu^*_F\,m_F}$. Solid line: $l=1$; dashed line:  
$l=3$; dot--dashed line: $l=5$. The condensate fraction is 
$n_B^0/n_B=0.6$, corresponding to a temperature of $70{\rm n\,K}$. 
The Bose--Fermi mixture is the same as in Fig.~\ref{fig1}.}
\end{figure}
In Fig.~\ref{fig3} the energy gap is displayed for three different 
values of $l$ at a fixed condensate fraction (~or temperature~). The fast 
decrease of the energy gap with increasing $l$ corresponds to the analogous 
behavior of the effective interaction. While the peak shown by 
$\Delta_l(k)$ about $k=k^*_F$ is due to the sharp increase of the 
effective interaction when the transferred momentum appoaches zero. 
This can be directly seen from  the approximate 
equation (\ref{pairl1}). 
\par
\begin{figure}
\includegraphics{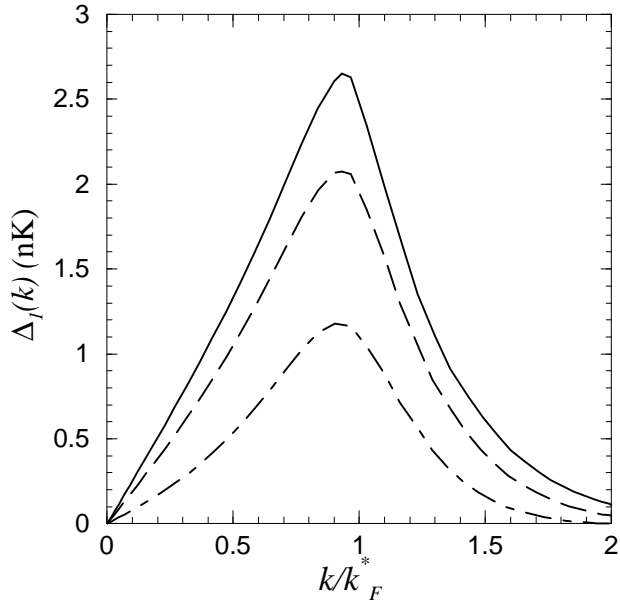}
\caption{\label{fig4}Energy gap in the $l=1$ channel for 
different values of the condensate fraction. Solid line: $n_B^0/n_B=0.4$ 
(~$T=87{\rm n\,K}$~); dashed line: $n_B^0/n_B=0.6$ (~$T=70{\rm n\,K}$~); 
dot--dashed line: $n_B^0/n_B=0.8$ (~$T=46{\rm n\,K}$~). 
The Bose--Fermi mixture is the same as in Fig.~\ref{fig1}.} 
\end{figure}

The behavior with the condensate fraction, of the energy gap for $l=1$ is 
reported in Fig.~\ref{fig4}. All $l$--channels show a similar 
behavior. We see that, when $n^0_B$ decreases, or, equivalently, 
the temperature increases, the gap $\Delta_1(k)$ increases for all values 
of $k$. This feature essentially reflects the analogous behavior 
of the effective--interaction strength. However, the fermion gas 
becomes less and less degenerate when the temperature increases, 
so that, the corresponding decrease of the effective Fermi momentum 
tends to counterbalance the effect of the effective interaction. 
Actually, explicit calculations show that, for $T\sim 100 {\rm nK}$ 
corresponding to $n^0_B/n_B\sim0.2$, the gap $\Delta_1(k)$ reverses 
its initial behavior. It should be observed that the approximation 
scheme proposed in this paper does not keep its validity for  
higher temperatures. In any case, the results of the present approach 
show that a superconducting phase in the $p$--wave channel can be 
achieved at temperatures $>100 {\rm nK}$. In Ref. \cite{Efr02} a 
critical temperature of $\sim 10^{-2} {\rm nK}$ was indicated for 
$p$--wave pairing in a mixture of $^6{\rm Li}$ and $^{87}{\rm Rb}$. For this 
mixture calculations performed within the present scheme and with the same 
values of parameters as in Ref. \cite{Efr02}, show that the behavior 
of the energy gap qualitatively does not differ from that of 
Fig.~\ref{fig4}. The deep discrepancy between the results of the present 
work and those of Ref. \cite{Efr02} should be ascribed   
to the fact that in the two approaches two opposite approximations 
have been used for the time dependence of the effective interaction. 
These approximations have already been discussed in Sec. IIB. 
\section{Summary and Conclusions}
In a mixture of ultracold Fermi and Bose gases, an interaction between 
fermions is induced by the exchange of elementary excitations 
of the Bose gas. Due to this supplementary interaction, a  
superconducting phase can appear in the Fermi component of the mixture.  
The mechanism is analogous to that accounting for superconductivity 
in ordinary metals. In this paper, a procedure to determine 
the induced interaction and to assess the possibility of a  phase 
transition has been proposed. For simplicity, a spatially homogeneous 
system has been considered. Both the gases have essentially been 
treated within a mean field approximation appropriate for the physical 
conditions realized in recent experiments: the values 
of temperature and densities are such that the Bogoliubov approximation 
for the Bose gas can be justified and the Fermi gas can be 
considered degenerate. Moreover, since the experiments concern gases 
confined by means of magnetic traps, in the present approach the 
atoms of both species have been considered completely polarized. 
This circumstance has allowed the bare fermion--fermion interaction 
to be neglected. 
\par
The same formalism, based on functional methods, has been used in all 
stages of the calculation. This allows us to introduce the various 
approximations to the effective interaction between fermions in a 
self--consistent way. In particular, the inclusion of the contributions 
from states containing two virtual phonons and the static approximation to 
its time dependence. This is more appropriate for the physical situations  
of actual experiments than the usual instantaneous approximation. 
Both terms of the interaction, with one and two virtual phonons, 
are attractive and of the same order of magnitude. They can give rise 
to the pairing of correlated couples of fermions with different values 
of the center of mass momentum. In this paper, only the pairing field 
with vanishing center of mass momentum has been considered, thus preserving 
the invariance of the system under space translations. Since the paired 
fermions lie in the same magnetic state, the pairing field is 
constrained to be an odd function of their relative momentum. 
Then, it contains only odd--$l$ components, when 
expanded in spherical harmonics. A particular feature of the 
corresponding gap equations is that they show solutions with 
only one non vanishing partial wave at a time. Then, different 
nonisotropic phases may occur in the Fermi component of the 
mixture, with the pairing fields decreasing their strength when 
their multipolarity increases. It is worth noting that the energy 
gaps increase with the temperature until a value of $T\sim 100 {\rm nK}$, 
corresponding to a high depletion of the Bose condensate $n^0_B/n_B\sim 0.2$. 
The approximations introduced in the present approach 
become less reliable for higher values of the temperature, for the 
values of boson and fermion densities assumed here. Nevertheless, we can 
expect that a superconducting phase can occur in a Bose--Fermi mixture 
at $T>100\,{\rm nK}$. This value is a relevant fraction of 
the Fermi temperature. 
\begin{acknowledgments}
I am grateful to M. Modugno, M. Baldo and H.-J. Schulze for valuable 
discussions. I would also like to thank A. Dellafiore for a careful 
reading of the manuscript. 
\end{acknowledgments}


\begin{thebibliography}{99}
\bibitem{Bar67} J. Bardeen, G. Baym, and D. Pines, Phys. Rev. {\bf 156}, 
207 (1967).
\bibitem{Tru01} A. G. Truscott, K. E. Strecker, W. I. McAlexander, 
G. B. Partridge, and R. G. Hulet, Science {\bf 291}, 2570 (2001).
\bibitem{Sch01} F. Schreck, L. Khaykovich, K. L. Corwin, G. Ferrari, 
T. Bourdel, J. Cubizolles, and C. Salomon, Phys. Rev. Lett. 
{\bf 87}, 080403 (2001). 
\bibitem{Had02} Z. Hadzibabic, C. A. Stan, K. Dieckmann, S. Gupta, M. W. 
Zwierlein, A. G\"orlitz, and W. Ketterle, Phys. Rev. Lett. {\bf 88}, 
160401 (2002). 
\bibitem{Roa02} G. Roati, F. Riboli, G. Modugno, and 
M. Inguscio, Phys. Rev. Lett. {\bf 89}, 150403 (2002); 
G. Modugno, G. Roati, F. Riboli, F. Ferlaino, R. J. Brecha, and 
M. Inguscio, Science {\bf 297}, 2240 (2002). 
\bibitem{Bij00} M. J. Bijlsma, B. A. Heringa, and 
H. T. C. Stoof, Phys. Rev. A {\bf 61} 053601 (2000). 
\bibitem{Hei00} H. Heiselberg, C. J. Pethick, H. Smith, and L. Viverit, 
Phys. Rev. Lett. {\bf 85} 2418 (2000). 
\bibitem{Efr02} D. V. Efremov and L. Viverit, Phys. Rev. B {\bf 65}, 
134519 (2002). 
\bibitem{Viv02} L. Viverit, Phys. Rev. A {\bf 66}, 023605 (2002).
\bibitem{Negele} J. W. Negele and H. Orland, {\it Quantum Many--Particle 
Systems} (Addison--Wesley, New York, 1988). 
\bibitem{Sto99} H. T. C. Stoof, in {\it Coherent atomic matter waves}, 
Proceedings of the Les Houches Summer School, Session 72, edited by 
R. Kaiser, C. Westbrook, and F. David 
(Springer--Verlag, Heidelberg, 2001), p. 219.
\bibitem{Walecka} A. L. Fetter and J. D. Walecka, {\it Quantum Theory 
of Many--Particle Systems} (McGraw--Hill, New York, 1971). 
\bibitem{Stra58} R. L. Stratonovich, Sov. Phys. Dokl. {\bf 2}, 416 (1958). 
\bibitem{Hub59} J. Hubbard, Phys. Rev. Lett. {\bf 3}, 77 (1959). 
\bibitem{Loff} A. I. Larkin and Yu. N. Ovchinnikov, Sov. Phys. JEPT 
{\bf 20}, 762 (1965); P. Fulde and R. A. Ferrel, Phys. Rev. {\bf 135}, 
A550 (1964).
\end{thebibliography}
\end{document}